\documentclass[preprint,journal]{vgtc}            

\onlineid{1155}

\preprinttext{To appear in IEEE Transactions on Visualization and Computer Graphics.}

\vgtccategory{Research}

\vgtcpapertype{Theoretical \& Empirical}

\title{Charting EDA: Characterizing Interactive Visualization Use in Computational Notebooks with a Mixed-Methods Formalism}
\author{%
  \authororcid{Dylan Wootton}{0000-0002-4453-6400}\and
  \authororcid{Amy Rae Fox}{0000-0003-0995-7899} \and
  \authororcid{Evan Peck}{0000-0001-9498-5362} \and
  \authororcid{Arvind Satyanarayan}{0000-0001-5564-635X}
}

\authorfooter{
  \item Dylan Wootton, Amy Rae Fox, and Arvind Satyanarayan are with MIT CSAIL. E-mails: \{dwootton, amyfox, arvindsatya\}@mit.edu, 
  \item Evan Peck is with University of Colorado Boulder. E-mail evan.peck@colorado.edu
}

\abstract{%
  Interactive visualizations are powerful tools for Exploratory Data Analysis (EDA), but how do they affect the observations analysts make about their data? We conducted a qualitative experiment with 13 professional data scientists analyzing two datasets with Jupyter notebooks, collecting a rich dataset of interaction traces and think-aloud utterances. By qualitatively coding participant utterances, we introduce a formalism that describes EDA as a sequence of analysis states, where each state is comprised of either a representation an analyst constructs (e.g., the output of a data frame, an interactive visualization, etc.) or an observation the analyst makes (e.g., about missing data, the relationship between variables, etc.). By applying our formalism to our dataset, we identify that interactive visualizations, on average, lead to earlier and more complex insights about relationships between dataset attributes compared to static visualizations. Moreover, by calculating metrics such as revisit count and representational diversity, we uncover that some representations serve more as "planning aids" during EDA rather than tools strictly for hypothesis-answering. 
We show how these measures help identify other patterns of analysis behavior, such as the "80-20 rule", where a small subset of representations drove the majority of observations. Based on these findings, we offer design guidelines for interactive exploratory analysis tooling and reflect on future directions for studying the role that visualizations play in EDA.

}


\teaser{
  \centering
    \includegraphics[width=\linewidth
    ]
    {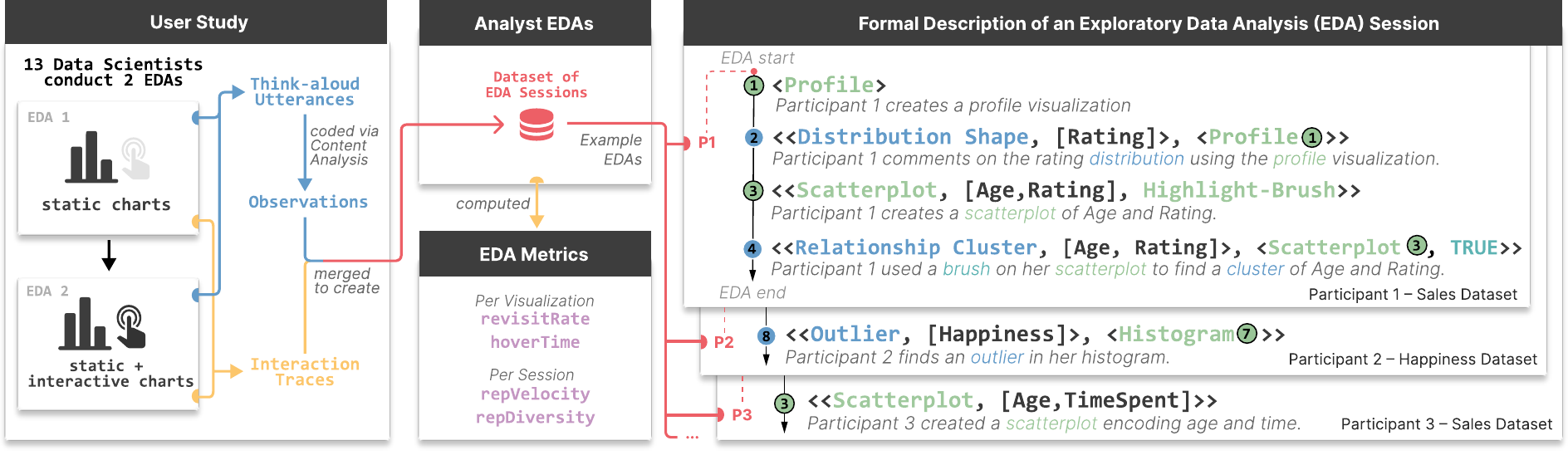}
    
    \caption{This study examined EDA practices via mixed methods. Think-aloud and interaction data from EDA sessions were collected and quantified using metrics and formal descriptions. The resulting dataset facilitated analysis of EDA behaviors and strategies.}
  
  \label{fig:teaser}
}

\graphicspath{{figs/}{figures/}{pictures/}{images/}{./}} 
\usepackage[svgnames]{xcolor} 
\usepackage{svg}
\usepackage{tikz}

\definecolor{oceanicred}{HTML}{EC5f67}
\definecolor{oceanicorange}{HTML}{F99157}
\definecolor{oceanicyellow}{HTML}{FAC863}
\definecolor{oceanicgreen}{HTML}{99C794}
\definecolor{oceanicteal}{HTML}{5FB3B3}
\definecolor{oceanicblue}{HTML}{6699CC}
\definecolor{oceanicpurple}{HTML}{C594C5}
\definecolor{oceanicbrown}{HTML}{AB7967}
\definecolor{annotationred}{HTML}{9B0000}
\definecolor{white}{HTML}{FFFFFF}

\newcommand{\fall}[1]{\texttt{\textcolor{oceanicred}{#1}}}
\newcommand{\foutput}[1]{\texttt{\textcolor{oceanicyellow}{#1}}}
\newcommand{\fvis}[1]{\texttt{\textcolor{oceanicgreen}{#1}}}
\newcommand{\fobs}[1]{\texttt{\textcolor{oceanicblue}{#1}}}
\newcommand{\freprusage}[1]{\texttt{\textcolor{oceanicteal}{#1}}}
\newcommand{\fmetric}[1]{\texttt{\textcolor{oceanicpurple}{#1}}}

\newcommand*\circledchart[1]{\tikz[baseline=(char.base)]{
    \node[shape=circle,draw,inner sep=1pt, fill=oceanicgreen, text=black] (char) {\textbf{\texttt{\footnotesize #1}}};}}
\newcommand*\circledobs[1]{\tikz[baseline=(char.base)]{
    \node[shape=circle,draw,inner sep=1pt, fill=oceanicblue, text=black] (char) {\textbf{\texttt{\footnotesize #1}}};}}
\newcommand*\circledrep[1]{\tikz[baseline=(char.base)]{
    \node[shape=circle,draw,inner sep=1pt, fill=oceanicyellow, text=black] (char) {\textbf{\texttt{\footnotesize #1}}};}}
\newcommand*\circledannotation[1]{\tikz[baseline=(char.base)]{
    \node[shape=circle,draw,inner sep=1pt, fill=annotationred, text=white] (char) {\textbf{\texttt{\footnotesize #1}}};}}  
    
\begin{document}
\maketitle

\section{Introduction}

The research literature widely considers interaction to play a central role in effective visualization for exploratory data analysis (EDA)~\cite{heer_interactive_2012, van_wijk_value_2005} because it supports a ``dialogue between the analyst and the data''~\cite{thomas_illuminating_2005}.
Recent empirical results, however, suggest a less clear picture. Studies have found no significant improvements in accuracy or error rates when using interactive visualizations for specific tasks such as bayesian reasoning or uncertainty communication~\cite{mosca_does_2021, yamamoto_ergonomic_2016}. Furthermore, a contextual inquiry with professional data scientists revealed that interactive visualizations are primarily used for communicating results rather than as a medium for conducting the analysis itself~\cite{batch_interactive_2018}. These findings suggest a gap between the theoretical benefits of interactive visualizations and their practical application in EDA.

We hypothesize two diagnoses for these discordant bodies of results. 
First, much of the work demonstrating the value of interactive visualization in EDA is conducted within systems purpose-built to support this activity (e.g., Tableau~\cite{battle_characterizing_2019}, Voyager~\cite{wongsuphasawat_voyager_2017}, VisTrails~\cite{bavoil_vistrails_2005}, among others~\cite{kale_evm_2023}). 
As a result, participants cannot ``opt out'' of the modality and conduct their analysis through other means (e.g., via code).  
%
Second, although existing approaches largely recognize that analysis is a \emph{situated} activity\,---\,that is, it involves human analysts working in a particular context, making observations with various representations of data\,---\,thus far, these methods often focus on one aspect of this behavior rather than synthesizing across it.
For instance, thematic analyses have been used to identify patterns of analytic behaviors~\cite{kale_evm_2023}, but it can be difficult to describe how these patterns manifest with particular interactive representations. 
On the other hand, quantitative approaches (e.g., interaction telemetry and provenance~\cite{nobre_revisit_2021,wongsuphasawat_voyager_2017}) capture detailed information about how analysts use particular representations. But without the context of qualitative insights, they can struggle to disambiguate observations. For instance, does hovering over a visualization indicate hesitation, gesticulation, or hypothesis testing?
Recent ``insight''-based approaches~\cite{battle_programmatic_2022} have come perhaps the closest to capturing the richness of analytic activity, but are presently focused on a narrow band of activity: quantitative insights described as data transformations.




To study how choices of data representation (including interactive and static visualizations) affect EDA, we aim to understand not only the \emph{what} of exploratory analysis (i.e., the insights gained) but also the \emph{how} (i.e., the evolving process and the use of different representations). 
To this end, we pose two research questions:
\vspace{-0.25em}
\begin{description}
\item[\textbf{RQ1:}] \label{RQ1} How do analysts' observations evolve over an EDA session?
\vspace{-0.25em}
\item[\textbf{RQ2:}] \label{RQ2} How do interactive and static data representations influence the processes and outcomes of EDA?
\end{description}

To address these questions, we conducted a qualitative experiment~\cite{QualExperiment2012} involving 13 data science professionals using Jupyter notebooks.
Participants were asked to complete two analysis tasks: the first with a lightweight library for authoring \emph{static} visualizations, followed by a second with an extended library including \emph{interactive} visualizations. 
Given their widespread use, Jupyter notebooks afford a more real-world context to study analytic behavior and, critically, do not \textit{presuppose} the value of interactive visualization.
Thus participants were free to forego visualization and interaction altogether, and simply author Python code using any third-party libraries they wished. 

To capture the full spectrum of analytic behavior, we recorded participants' verbal utterances and telemetry, merging these data streams through a content analysis~\cite{Hsieh2005_contentAnalysis} to create a \textit{unified dataset of analytic activity}. To analyze this dataset, we developed a novel \textit{formalism} that models EDA sessions as a sequence of analysis states.
Each analysis state is either the representation an analyst constructed (e.g., the output of a dataframe, or an interactive visualization), or an observation they made (i.e., an utterance about one or more representations).

To address RQ1, we leverage our formalism to code and track analyst observations over time. We identify 15 distinct types of utterances, grouped into four categories: utterances about \emph{dataset} size or orientation, or whether there was any missing data; utterances about \emph{variable} distribution or outliers; \emph{relationship} utterances that expressed concepts including strength, directionality, and clustering; and \emph{process} utterances that described intended analysis steps, or meta characteristics about a representation. Our analysis of these observations shows they follow distinct temporal patterns during EDA (\S~\ref{fig:utterances-over-time}). 
Analysts tend to address dataset-level metadata early on, while variable distributions and relationship insights occur throughout the analysis.
Notably, interactive visualization accelerate relationship utterances, with these statements occurring 15\% earlier than under the static condition.

To investigate RQ2, we leverage our formalism to combine representational telemetry with analyst observations, enabling us to explore the co-occurrence of representation use and analytical insights. We introduce a series of quantitative metrics including \emph{revisit count}, or the total number of times a participant hovered over a representation; \emph{output velocity}, or the number of representation instances created per unit time; and, \emph{representational diversity}, or the number of unique representation types created during an analysis.
We use these metrics to investigate patterns of exploration, revealing how some participants achieved broad coverage during their EDA (\S~\ref{broad-exploration}). 
Furthermore our formalism uncovers patterns in representation usage. Notably, we observe an 80-20 rule of representation use (\S~\ref{RESULTS-80-20}) and the propensity to use all-attribute representations as aids to plan analyses (\S~\ref{sec:planning-aids}). Taken together, our work contributes to calls for "deepening [the] theoretical foundation" of exploratory data analysis \cite{hullman_designing_2021}. 

\section{Related Work}

Our work continues a tradition of studying EDA through technical and empirical approaches. In these section, we review these prior studies\,---\,organized by their methodological choices\,---\,and contrast their results with our objectives. 

\textbf{Attribute Methods:}
Attribute-based methods have provided valuable insights into how analysts explore data features during EDA. These approaches operationalize EDA by quantifying the number and combinations of attributes that analysts examine, using metrics such as attribute-set counts \cite{wongsuphasawat_voyager_2017,sarvghad_visualizing_2017,alam_analyzing_2017} or search trees structure \cite{battle_characterizing_2019}. These metrics facilitate comparing different analysis sessions, enabling researchers to assess how various interventions affect the breadth and depth of attribute exploration during EDA. Moreover, they reveal structural elements of the exploration process. For instance, Battle \& Heer's study of analysts using Tableau identified key ``analysis-states''\,---\,particular attribute combinations that played pivotal roles in participants' explorations \cite{battle_characterizing_2019}. Notably, their study finds that analysts using Tableau often prefer depth-oriented exploration, thoroughly investigating specific attribute relationships, rather than employing a breadth-oriented approach that surveys a wide range of different attribute sets.
Our work extends these results by describing how particular representations shape attribute exploration. For example, we find analysts engage in \textit{attribute addition} when using interactive visualizations (\S\ref{attribute-addition}) alongside other strategies used to broadly cover data attributes (\S\ref{broad-exploration}).




\textbf{Insight Methods:} Insight methods focus on identifying and characterizing the analytical knowledge generated during EDA \cite{north_toward_2006}. These methods typically employ think-aloud processes \cite{north_toward_2006, boggust_embedding_2022} or elicit insights through open-ended responses \cite{nobre_evaluating_2020}.
Researchers then code these insights based on their semantic content, such as \textit{Generalization} or \textit{Hypothesis} \cite{liu_effects_2014}, and analyze additional qualities like whether insights are broadening or deepening \cite{sarvghad_visualizing_2017} or their factual correctness \cite{zgraggen_investigating_2018}.  These coded utterances are often aggregated to compute metrics like time-to-first insight and total number of insights \cite{boggust_embedding_2022,liu_effects_2014,zgraggen_investigating_2018,north_toward_2006, guo_case_2016}.

We differentiate our approach from previous insight methods through the use of qualitative content analysis to record both \textit{what is said} and \textit{what representations were used} to make such utterance. 
By explicitly linking the insight to the representation, our work investigates how different representations co-occur with particular insights. As a result, we compute aggregated information about insights during analysis conditions (\S\ref{RESULTS-observations}) but also investigate how insights are formed using particular representations (\S\ref{representation-overview}). This approach lets us understand the impact of visualizations on the EDA process, such as analysts deriving 80\% of their insights from just 20\% of their representations (\S\ref{RESULTS-80-20}). Furthermore, our qualitative content analysis captures a wider range of insights, demonstrating how specific visualizations correspond to particular types of observations (Fig. \ref{fig:output_creation_and_use}).

\textbf{Interaction Traces:}
Interaction traces provide rich quantitative data to describe analyst activity, offering insights into specific measurable behaviors during EDA. These traces range from simple actions like chart hovers \cite{sarvghad_visualizing_2017,wongsuphasawat_voyager_2017} to complex action sequences within interactive visualizations \cite{nobre_revisit_2021}. Researchers have leveraged these logs to create metrics assessing exploratory behavior and to reveal how user characteristics influence exploration patterns \cite{feng_patterns_2019}.
However, a key limitation of interaction traces is their inability to capture the meaning behind interactions. A hover over a chart could represent an insight being made or analyst confusion. To address this, researchers often combine interaction traces with other characterization strategies. In attribute-based methods, for example, they help demonstrate when a particular set of attributes is "considered," from hovering over visualizations \cite{wongsuphasawat_voyager_2017} to creating them in Tableau \cite{battle_characterizing_2019}.
In our work, we link interaction traces to utterances, revealing how specific interaction patterns can indicate different analysis strategies. For instance, we calculate a revisit count for each representation based on hover frequency, and used this metric to identify that a subset of highly revisited charts are frequently associated with analysis planning behaviors (\S\ref{sec:planning-aids}). 

\textbf{Modeling Notebook Corpora:}
Recent research has explored modeling notebooks and their histories, primarily focusing on predicting future analyst actions given the current notebook state. For instance, Auto-Suggest \cite{yan_auto-suggest_2020} uses a recurrent neural network trained on notebook corpora to generate future data transformation operations. Similarly, EDA Assistant \cite{li_edassistant_2023} ranks slices of programs from similar notebooks and provides frequently used next steps. Other approaches have focused on generating entire EDA sessions rather than snippets of code. For example, Bar et al. \cite{bar_el_automatically_2020} formulate EDA as a control problem where they use a reward signal based on the novelty and diversity of insights to automatically generate entire EDA sessions.
While these systems develop useful tools to facilitate EDA, they primarily aim to predict the analyst's next action rather than providing insights into broader patterns of analytic behavior during EDA. In contrast, our work seeks to understand the cognitive processes and decision-making patterns that underlie analysts' interactions. Future systems-building work could use the results of our analyses to better model analyst activity and recommend next steps.

\textbf{Interviews and Surveys:} 
Interview and survey studies provide crucial insights into the real-world practices of data scientists, shaping our understanding of EDA workflows. Kandel et al. conducted foundational work understanding the stages of data science work \cite{kandel_enterprise_2012}. They interviewed data scientists across various enterprise organizations outlining five key job responsibilities: \textit{discovery}, \textit{profiling}, \textit{data wrangling}, \textit{modeling}, and \textit{reporting}. These elements are central to data science activities. Further refining this understanding, Wongsuphasawat et al. conducted interviews that revealed a more detailed set of 16 analytic behaviors, such as \textit{converting data formats} and \textit{examining bivariate plots} \cite{wongsuphasawat_goals_2019}. Interviews also enable researchers to investigate attitudes towards particular EDA tools, such as Batch et al.'s \cite{batch_interactive_2018} work to understand the ``Interactive Visualization Gap'' in EDA. Furthermore, when conducting empirical studies, surveys are often administered following an exploratory analysis session \cite{gadhave_persist_2023,wu_b2_2020, sarvghad_visualizing_2017}. Most commonly,  surveys include questionnaires like the NASA-TLX\cite{hart1988development} for understanding subjective workload during a task \cite{gadhave_persist_2023} or Likert scale questions to elicit preferences when using a tool \cite{wu_b2_2020,sarvghad_visualizing_2017}. 
Our work builds on these findings by examining how analysts use different representations during EDA (\S\ref{representation-overview}), providing a more nuanced understanding of \emph{when} and \emph{why} certain visualizations are used\,---\,an approach that allows us to bridge the gap between reported practices and actual behavior.

\textbf{Thematic Analysis:} Thematic analysis seeks to identify occurrences of broad behavioral patterns or \textit{themes}\cite{pu_how_2023, kale_evm_2023, boggust_embedding_2022}. These approaches typically involve participants thinking aloud in order interpret the meanings of behaviors given their context. For example, Kale et al. \cite{kale_evm_2023} investigated the effect of a tool that enables \textit{model-checking} through a within subjects comparison. Using thematic analysis, they characterize how the patterns of analysis shifted when the model-checking functionality was introduced, revealing that this tool "structure[d] participants’ thinking around one or two long chains of operations". In contrast to our study, thematic analysis does not seek to characterize the content of entire analysis session, choosing instead to focus on larger themes that were observed during exploration.

\section{Methods}
\label{METHODS}

Our research questions aim to describe the temporal progression of analysts' observations and inferences (\textbf{RQ1}), while also comparing how these behaviors unfold with \textit{static} vs. \textit{interactive} visualizations (\textbf{RQ2}). 
These research questions are both \textit{descriptive} and \textit{comparative} in nature. To address these questions comprehensively, we adopted a hybrid design that combines task observation and semi-structured interviews within the framework of a repeated-measures experiment. This approach, described in the mixed methods literature as a \textit{qualitative experiment} \cite{QualExperiment2012}, allows us to capture rich, contextual data about analysts' thought processes and actions while also enabling systematic comparisons between static and interactive visualization conditions.

\subsection{Study Design, Procedure, and Participants}
\label{METHODS-study-design}


Our independent variable is \textbf{representation interactivity} with two levels: \textit{static} and \textit{interactive}. We use a repeated-measures (i.e. within-subjects) structure where we measure participant behavior in two tasks (static, interactive), and with two datasets that are counterbalanced in their assignment across the two tasks. Note that we \textit{did not} counterbalance static/interactive task order because the interactive features necessarily built upon knowledge of the static visualizations. Participants engaged in a 90-minute video-conference divided into four parts: introductions/informed consent, two \textit{EDA sessions}, and an interview.

Each \textit{EDA Session} began with an introduction to the (static/interactive) features of the visualization library (\textit{Features Intro}), followed by an opportunity for the participant to explore the new APIs via sample code (\textit{Features Tutorial}). Next, participants were given a notebook with a dataset and scenario for an \textit{Analysis Task}, and asked to complete an exploratory analysis in approximately 25 minutes while thinking aloud. Throughout this process, their interactions with the notebook – running code cells, brushing on charts, and scrolling – were recorded as interaction telemetry. The structure of the static task was identical, with the dataset counterbalanced across participants. Each session concluded with a semi-structured interview and debrief.

We recruited 16 participants through social media, personal networks, and crowdwork platforms. 
Two participants were involved in pilot studies to refine data collection procedures. Of the 16 participants who completed the study, three were excluded due to either incomprehensible think-aloud responses or an insufficient level of Python proficiency.
Our resultant pool comprised 13 participants: 4 women, 8 men, and one person who identified as non-binary; participant ages ranged between 27 and 41 years (average age 31). 
All participants regularly conducted EDA using Jupyter notebooks as part of their occupation. 
Their most common job title was Data Scientist (5), followed by PhD Candidate (3), Software Developer (2), Data Analyst (1), Economist (1), and Statistician (1).





\subsection{Controlling for Library Expertise with Altair Express}
\label{ALX-design}

To facilitate comparisons between participants' behaviors, it was essential that they used the same visualization library.
However, this introduces a confound: participants' existing expertise with visualization packages. 
To control for this, we developed a new visualization package to establish a common baseline of relative novelty for all participants. 

Our library, called \textit{Altair Express (ALX)},\footnote{The name was chosen to mirror the relationship between Plotly and Plotly Express. That is, \textit{Altair : Altair Express :: Plotly : Plotly Express}.} is a Python-based visualization package that offers a high-level declarative API for specifying interactive visualizations. 
In contrast to the composable approach of the existing Altair visualization package (and its underlying grammar Vega-Lite~\cite{satyanarayan_vega-lite_nodate}), ALX instead provides a \emph{typology} of visualizations and interaction techniques\,---\,an approach we chose to reduce specification friction analysts might face during EDA. 
We surveyed existing Python-based chart typologies (e.g., Plotly Express, Seaborn, etc.) and implemented the set of statistical charts we hypothesized to be most relevant to EDA including: \verb|barplot|, \verb|countplot|, \verb|hist|, \verb|jointplot|, \verb|lineplot|, \verb|heatmap|, \verb|pairplot|, \verb|profile|, \verb|scatterplot|, and \verb|stripplot|.

The interaction typology in ALX is defined by \textit{effect-action} pairs: an \textit{effect} is the change that occurs when a user performs an interaction (e.g., showing a tooltip, zooming into a region, etc.), and an \textit{action} is the event that triggers the interaction (e.g., clicking, brushing, etc.). Thus, the typology comprises: \verb|highlight_brush|, \verb|filter_brush|, \verb|tooltip_hover|, \verb|pan_zoom|, \verb|filter_slider|, \verb|filter_type|, \verb|highlight_color|, and \verb|highlight_point|. 





Using the \verb|+| operator, visualization and interaction types can be composed together. For instance, \texttt{alx.highlight\_brush() + alx.scatterplot(data, x='Weight', y='Horsepower')} produces a scatterplot of the \texttt{Weight} and \texttt{Horsepower} of cars; users can brush the scatterplot highlighting selected points in blue and dimming the rest to gray. 
Using \verb|+|, users can add multiple interaction techniques to a single visualization, or concatenate multiple static and/or interactive visualizations together to produce a custom dashboard. 
ALX implements these interactive visualizations via Vega-Lite~\cite{satyanarayan_vega-lite_nodate}. 





Finally, in addition to its specification language, ALX implements a handful of features designed to address limitations researchers have identified of using interactive visualizations in computational notebooks~\cite{wu_b2_2020, batch_interactive_2018}. 
For example, with ALX, analysts can use a ``\textit{copy-and-paste}'' in order to extract an underlying data selection. When a selection is made\,---\,for instance, by clicking on a point, dragging a slider, or brushing\,---\,the analyst can press \verb|control + c| to copy the pandas query necessary to select the data. This query can then be pasted into the subsequent cells in the notebook to filter down to the selected data for further investigation or charting.
\vspace{-0.5em}
\subsection{Data Analysis Procedure}
\label{METHODS-codebook-development}

We applied an inductive content analysis ~\cite{Hsieh2005_contentAnalysis, Marsh2006} to the rich stream of video and think-aloud data our participants produced. 
We split transcripts of the video recordings into discretized units of meaning we call \textit{utterances}.
And, using participants' screenshare, mouse gestures, and linguistic prosody, we coded what representations participants used in the process of making a particular utterance. 
We limited the scope of our coding to only include the \textit{Analysis Tasks}\,---\,thus, we excluded utterances participants made when they were familiarizing themselves with ALX's features, debugging, or during the post-interview.


The first and second authors followed an inductive process consistent with the application of grounded theory in HCI \cite{WOK-groundedTheory, Hsieh2005_contentAnalysis} to develop a codebook for categorizing participants' utterances. 
This processes involved eight iterations of independent coding centered on: (1) developing structure, (2) aligning criteria, and (3) reconciling discrepancies. In the final round of reconciliation, the first and second authors independently coded a random sample of 100 utterances, to calculate an Inter-Rater Reliability (IRR) measure of Krippendorf's $\alpha = 0.85$.\footnote{Krippendorf's alpha is the recommended IRR metric for multi-code structures where more than one can can be applied to one observation. Using a more generous alternative we calculate reliability of (\textit{Observed Agreement}=0.87). In both cases our IRR passes normative thresholds of reliability \cite{Lombard2004}.} 

\section{A Formal Description of EDA Sessions}

\begin{figure}[!t]

  \includegraphics[width=\columnwidth]{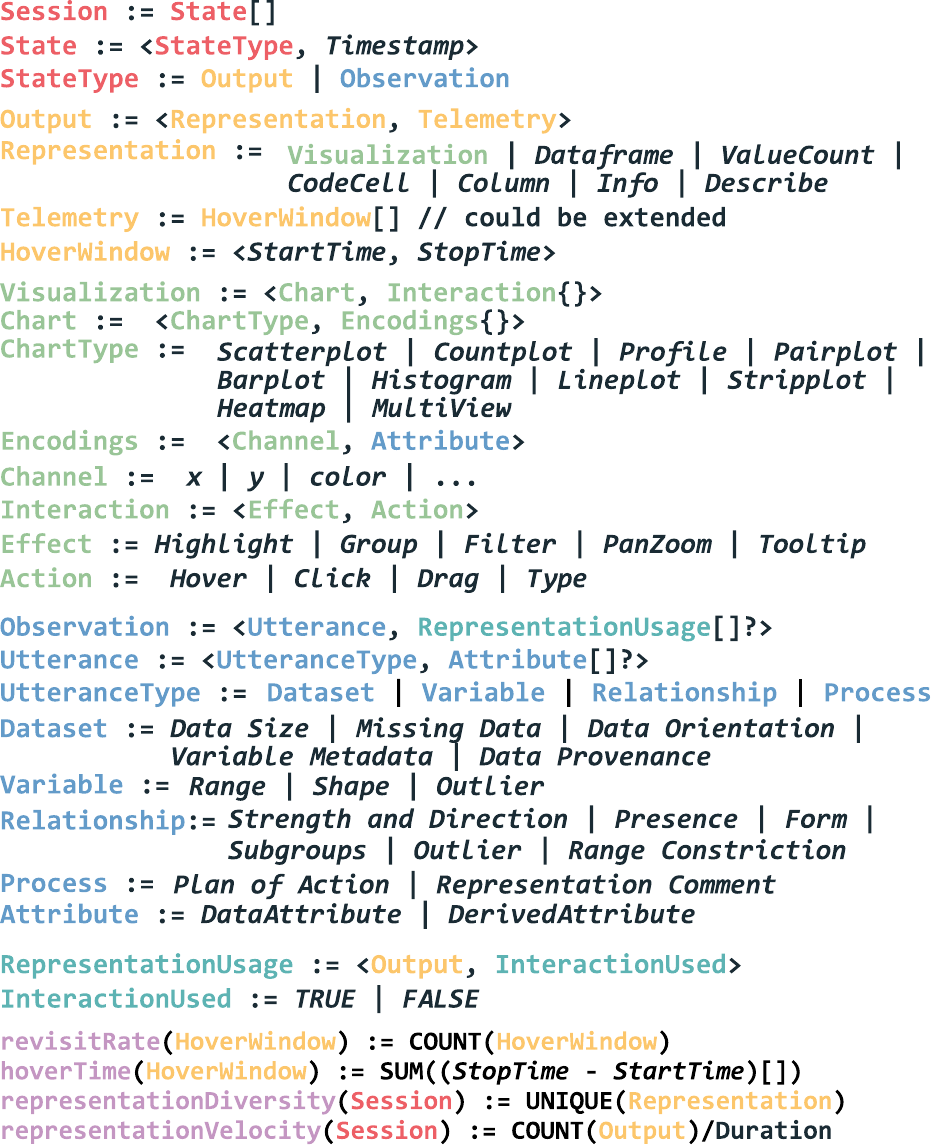}
  \caption{A formal definition of EDA sessions in terms of analysis states that comprise either a representation alone (e.g., a visualization, dataframe output, etc.) or an observation made with one or more representations. Italics indicates terminal symbols.}
  \label{fig:formalism}
  \vspace{-2.5ex}

\end{figure}

\begin{figure}[!t]
  \includegraphics[width=\columnwidth]{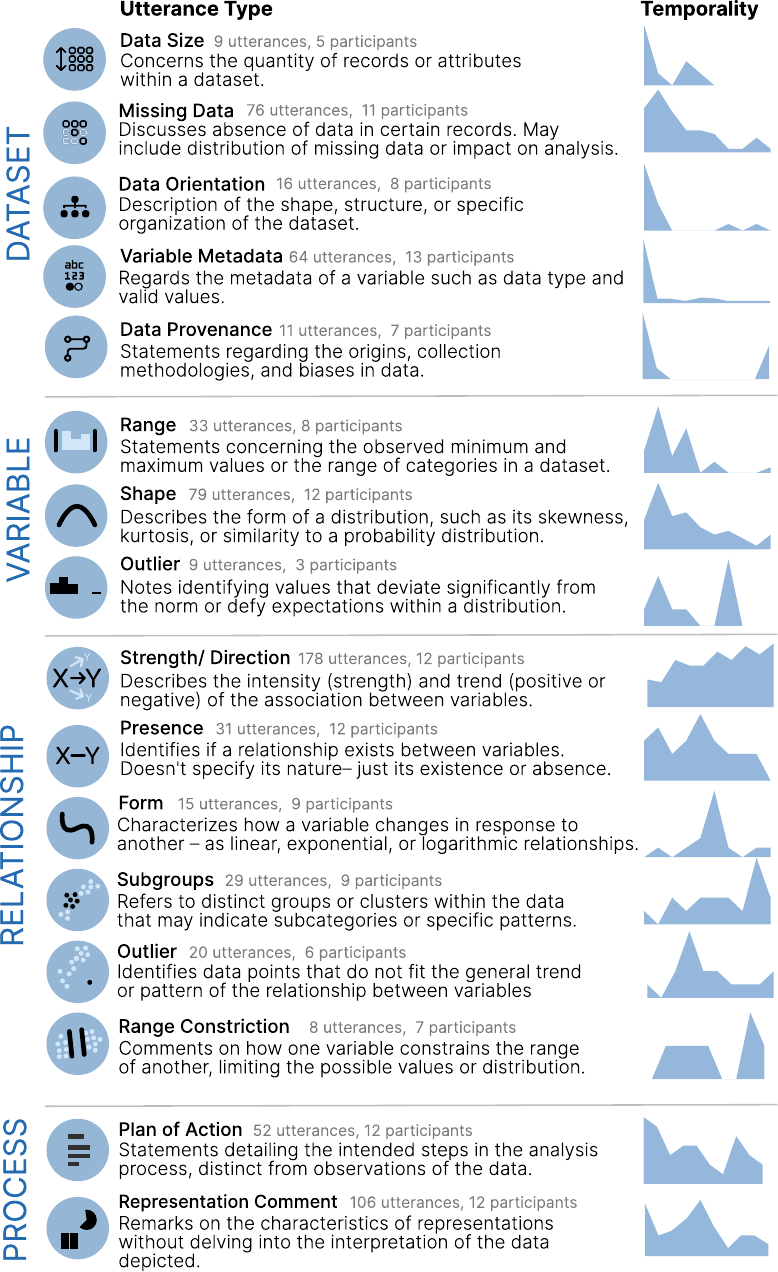}
  \caption{Utterances are structured as a 2-level hierarchy, with the highest level codes (\emph{Dataset}, \emph{Variable}, \emph{Relationship}, \emph{Process}) describing the general topic of an utterance, and lower level detail codes delineating the utterance's content more precisely.}
  \label{fig:highlevelutterances}
\vspace{-4ex} 

\end{figure}

We express the results of our mixed-methods analysis through the formal description shown in Figure~\ref{fig:formalism}.
We find an EDA \fall{Session} progresses through a sequence of analysis \fall{States}. 
Each \fall{State} can either be a standalone \foutput{Representation} (e.g., a \fvis{visualization}, dataframe printout, etc.) or be a verbal \fobs{Observation} that an analyst makes.
For each representation, we collect a variety of \foutput{Telemetry} data, but our analysis focuses only on \foutput{HoverWindows} (i.e., time spans of when a participant hovered over a given representation)\,---\,we leave other abstractions that can be derived from telemetry data to future work. 

\fobs{Observations} associate verbal \fobs{Utterances} with any  \foutput{Representations} used to make them, as indicated through \freprusage{RepresentationUsage}. We distinguish these observations into those made with interactive features (such as brushing or tooltips, coded as \freprusage{InteractionUsed}) from those on interactive charts that did not utilize interactivity. 
We use the term \fobs{Utterance} rather than \emph{insight} or \emph{inference} to indicate that, even with the context of the participant's screenshare, mouse gestures, and linguistic prosody, we cannot precisely determine the participant's state of knowledge.
Thus, we work to interpret as much of each utterance's semantic content as possible via our qualitative coding procedure.

As Figure~\ref{fig:highlevelutterances} shows, this procedure yielded 16 \fobs{UtteranceTypes} spread across four categories: utterances about the overall \fobs{Dataset} including its size, orientation, quality, provenance, and metadata; utterances about individual \fobs{Variables} including about the distribution of data values (e.g., min, max, outliers) and the shape of this distribution; utterances about \fobs{Relationships} between variables including whether any relationship exists and, if so, what form, strength, and direction this relationship takes; and, finally, utterances about the overall analytic \fobs{Process} including statements about intended next steps or remarks about representations that are not about depicted data.

We find this formalism offers unique insights into EDA activity, illustrated by the following vignette inspired by participant behavior:

\newenvironment{myquote}%
  {\list{}{\leftmargin=0.1in\rightmargin=0.1in}\item[]}%
  {\endlist}
  
\begin{myquote}

Ada, a data analyst at an e-commerce company, is tasked with investigating a customer purchase behavior dataset that includes \emph{customer age}, \emph{product categories}, \emph{shipping speed}, and \emph{customer satisfaction ratings}.
Ada begins by creating a data profile \circledchart{1}, a multiview visualization with concatenated univariate histograms for each variable. While examining the distributions, she notices an unusual pattern in the \emph{satisfaction ratings} \circledobs{2} - there's a concerning spike at 1-star ratings, contrary to the company's belief that customer satisfaction was generally high. Intrigued, Ada uses a crossfilter interaction to brush over the 1-star ratings, and observes a shift in the age distribution in the profile, noting that these dissatisfied customers tend to be younger \circledobs{3}.  

To investigate further, Ada creates a scatterplot of \emph{satisfaction ratings} vs. \emph{customer age} \circledchart{4}. The scatterplot confirms a cluster of younger customers with low satisfaction ratings \circledobs{5}. Ada isolates this cluster using a brush selection tool and examines the associated customer details in a table view \circledrep{6}. 
Digging deeper into the table, Ada discovers that a significant portion of these customers' purchases are from the "Gifts" category, and their \emph{shipping speed} is often listed as "expedited" \circledobs{7}, suggesting young buyers might be using the platform primarily for last-minute gift purchases, resulting in higher stress and lower satisfaction when issues arise.

  \vspace{-2ex}
\end{myquote}
\vspace{-0.5em}
\hfill \break
Using attribute-based metrics~\cite{wongsuphasawat_voyager_2017, battle_characterizing_2019}, we might view Ada's EDA as a three-step process: analyzing \emph{all} attributes (with the profile); then analyzing \emph{age} and \emph{rating}; and finally, returning to \emph{all} attributes (with the data table).
This approach makes it difficult to identify that Ada did not ever actually analyze particular attributes (e.g., \emph{purchase history}) despite their inclusion in certain representations (i.e., the profile and data table).
Moreover, by being representation-agnostic, attribute-centric metrics treat the profile and data table as equivalent and, as a result, miss the different ways Ada used these two views\,---\,for instance, that she brushed the profile view to reveal a relationship between \emph{age} and \emph{satisfaction} versus examining the table in a more record-by-record fashion. 
These issues are compounded when applying attribute-centric metrics to analyze interactive visualization as the space of possible observations is greatly expanded~\cite{jun_hypothesis_2022}.

Task and insight-based methods often do not account for representation either.
As a result, they ignore analytic expressions that are not verbalized and instead latently conveyed via the representation\,---\,that is, the act of making a chart is intrinsically an inquiry, even if it is not used to make an observation out loud. 
Moreover, depending on the granularity of task/insight codes, these methods may miss important nuance in Ada's activity. 
For instance, with the protocol followed by Zgraggen et al.~\cite{zgraggen_investigating_2018}, one might label Ada's analysis as a \emph{Distribution Shape} insight followed by two \emph{Correlation} insights\,---\,a strategy that collapses insights about ``clusters'' and ``correlations'' together.
More recent insight-based approaches, such as the formalism developed by Battle \& Ottley~\cite{battle_programmatic_2022}, begin to address many of these shortcomings\,---\,for instance, they formalize an \texttt{AnalyticKnowledgeNode} to encompass data relationships and transformations. 
While this method would be able to capture much of Ada's activity (e.g., interactive brushing as issuing a series of data queries), it is focused only on describing the \emph{quantitative insights} a participant might make about a dataset. 

\begin{figure}[t!]
  \centering
  \includegraphics[width=\linewidth]{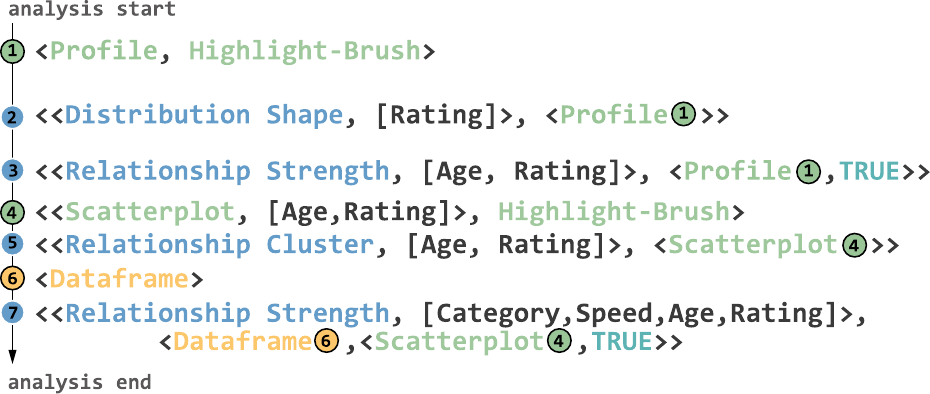}
  \caption{Example of Ada's analysis session encoded in our formalism. For clarity, we have omitted some levels of nesting for the formal description of this example. Colors are associated with the corresponding formalism construct: \foutput{Output} (non-visualization outputs), \fvis{Visualizations}, \fobs{Observations}, and \freprusage{Representation Usage}.}
  \label{fig:formalism-example}
   \vspace{-1.5em}
\end{figure}

In contrast, our formalism separately records the representations Ada constructed, the utterances she verbalized, and links the two together as a series of observations (Fig.~\ref{fig:formalism-example}). This description better reflects the situated nature of EDA\,---\,that observations occur \emph{with} representations, and that non-verbalized representations can play important roles in an analysis session. In the subsequent sections, we demonstrate how to apply the formalism to investigate behaviors during EDA.


\section{Characterizing Analyst Utterances}
\label{RESULTS-observations}

In this section, we analyze participant \fobs{Observations} to investigate the semantic content of analyst EDAs and how they evolve over time (\textbf{RQ1}). We examine the temporal patterns of different types of observations throughout EDA sessions (\S\ref{sec:temporal-patterns}), comparing how these patterns manifest in static versus interactive conditions. Additionally, we explore the transitions between different types of observations, extending our understanding of exploratory behaviors beyond the previously identified touring motifs \cite{kale_evm_2023} (\S\ref{RESULTS-Sequential-transitions}). This analysis provides insights into the structure of EDA processes and how they are influenced by the availability of interactive visualizations.

\label{CA-UTTERANCES}
\subsection{Temporal Patterns}
\label{sec:temporal-patterns}


\begin{figure}[t!]
  \centering
  \includegraphics[width=\columnwidth]{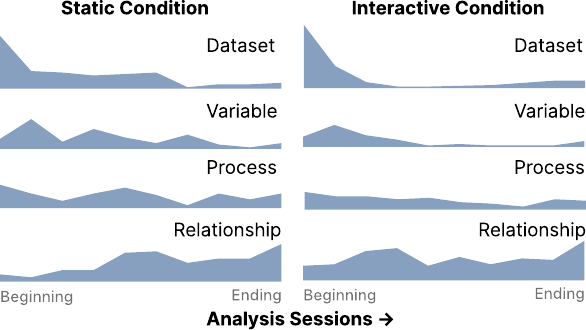}
  \caption{Occurrence of utterances categories throughout analyses.}
  \label{fig:utterances-over-time}
  \vspace{-1.5em}
\end{figure}

As the area charts in Figure~\ref{fig:utterances-over-time} show, we find that while  analysts' processes align \textit{in aggregate} with traditional, linear EDA models (from individual variable analysis and then relationship exploration \cite{kandel_enterprise_2012}), the analysis process is both more fluid and sensitive to interactivity than rigid interpretations of those models would suggest. To examine analyst processes, we calculated the median moment through the analysis session (expressed as a percentage) in which analysts made \fobs{Observations} across our four \fobs{UtteranceTypes}: \fobs{Dataset} (13.43\%), \fobs{Variable} (25.60\%), \fobs{Relationship} (56.86\%), and \fobs{Process} (40.18\%).


In particular, interactive EDA sessions prompted earlier observations about \fobs{Relationships} in the data (IQR 28\%–75\% through a session) compared to static EDAs (IQR 43\%–85\%). We hypothesize that the use of interactive profiles, featuring cross-filterable univariate histograms, encouraged analysts to explore relationships sooner. Our subsequent findings of analysts switching from static to interactive profilers (\S~\ref{attribute-addition}) support this: many participants shifted from \fobs{Variable}  to \fobs{Relationship} utterances almost immediately upon encountering the interactive profile. This finding opens questions about whether the affordances (or \textit{presence}) of interactive profiles enables bypassing distribution analysis, and whether we can articulate the tradeoffs of such process changes.
More broadly, the presence of relationship utterances across both static and interactive EDA sessions suggests that analysts are willing, perhaps even eager, to explore \fobs{Relationships} before fully developing a mental model of individual \fobs{Variables}.


    
  
\subsection{Sequential Transitions}
\label{RESULTS-Sequential-transitions}
\begin{figure}[t!]
  \centering
  \includegraphics[width=\linewidth,alt={This image presents a detailed analysis of utterance transitions during Exploratory Data Analysis (EDA) sessions. The left side shows a transition matrix of sequential utterances, where the size of each circle represents the frequency of transitions between different utterance types. The matrix reveals patterns in how analysts move between different types of observations during their analysis.
On the right side, the image highlights a "Variable Gap" in transitions between Interactive and Static analyses on the happiness dataset. This comparison shows how the use of interactive tools affects the sequence of observations, particularly in relation to variable-related utterances.}]{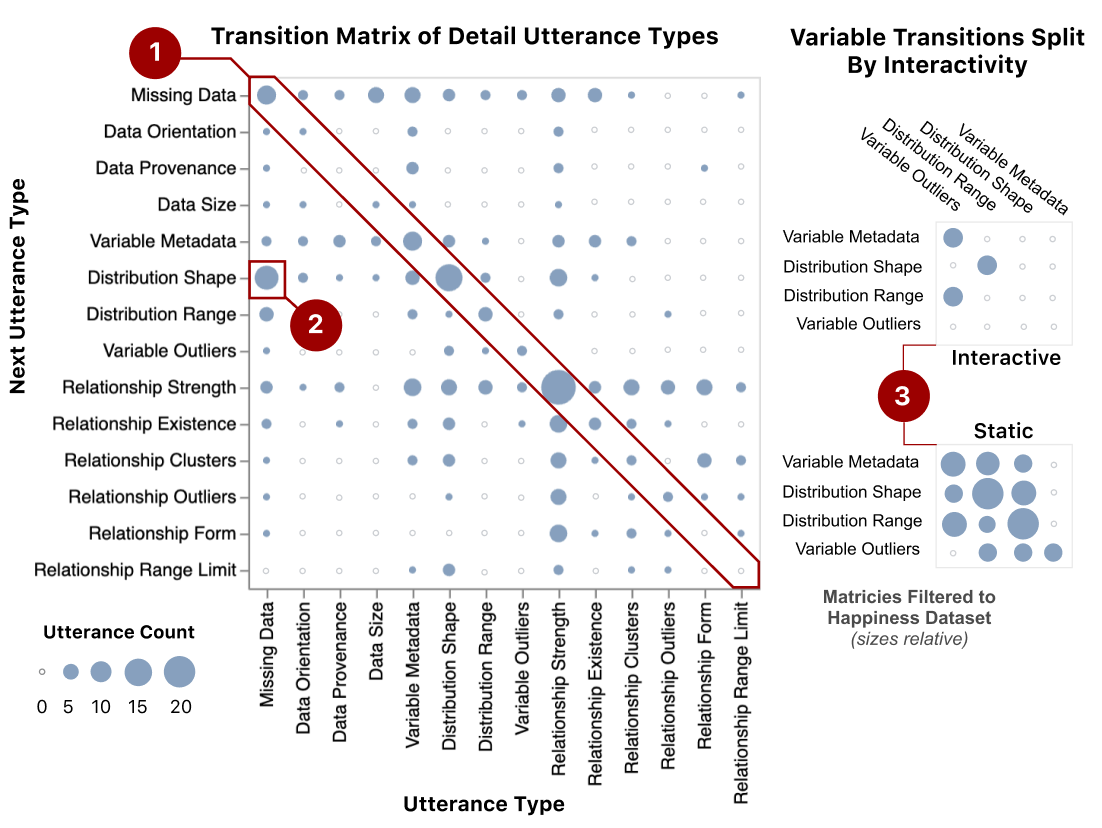}
  \caption{(left) A transition matrix of sequential utterances. (right) The transition matrices showing the "Variable Gap" in transitions between Interactive and Static analyses on the happiness dataset.
  }
  \vspace{-2em}
  \label{fig:sequential-utterances}
\end{figure}


During their analyses, participants made seven different types of utterances on average. Looking at the sequential transitions between utterances reveals a number of common analysis motifs \cite{kale_evm_2023}.

 
\textbf{Tour-Driven Exploration} Fig~\ref{fig:sequential-utterances}~\circledannotation{1}: Frequent self-transitions between similar utterance types (e.g., multiple consecutive utterances focused on \fobs{Relationship} strength) suggest that analysts often adopt a systematic ``touring'' approach during EDA. This finding aligns with concepts of univariate and bivariate tours \cite{kale_evm_2023,  lee_lux_2021}, where analysts methodically explore specific aspects of individual \fobs{Variables} and their \fobs{Relationships}. However, we observed self-transitions extending beyond \fobs{Relationship} analysis to include utterances about \texttt{Missing Data} and \texttt{Variable Metadata}. This suggests that ``touring'' behaviors are broader than previously described \cite{kale_evm_2023}.




\textbf{Column- vs. Row-Centric Missingness} Fig~\ref{fig:sequential-utterances}~\circledannotation{2}: The most common transition between utterance types was moving from \texttt{Missing Data} to \texttt{Distribution Shape}. This often occurred early on in analyses through use of profile visualizations. The design of profile presents missing data alongside the column's distribution, subtly promoting a column-centric view of missingness. However, as a counter-example, P10 investigated missingness as a characteristic of individual data records (rows), skipping the profile entirely. Visualizing the missingness per record on a scatterplot, he commented \textit{``... most of the rows have no missing columns, and then they progressively have more and more. So I guess, depending on what the analysis we're gonna do is, we may or may not exclude data points.''} 
This approach highlights different potential causes for missingness and raises a design question:  how can profile encourage analysis of column- and row-level missingness?


\textbf{The ``Variable Gap'' and Interactive Profiles} 
\label{sec:variable-gap}
Fig~\ref{fig:sequential-utterances}~\circledannotation{3}: In the happiness dataset, many participants skipped characterizing \fobs{Variables} altogether, instead immediately focusing on \fobs{Relationships}. This caused a \textit{Variable Gap} between conditions, visible in the transition matrices (right). This shift often coincided with the use of an interactive profile—a tool comprising univariate distribution visualizations that supports cross-filtering. For example, participant P5 initially followed a variable-first pattern in her static analysis, narrating out 6 distributional utterances about her variables using the profile. Upon beginning her interactive analysis, she immediately began making relationship utterances by cross filtering on the profile view (see \S~\ref{attribute-addition} for more information). 
\section{Characterizing Representations and Usage}
\label{RESULTS-rep-usage}

Guided by \textbf{RQ2}, we explore the link between \foutput{Representations} and \fobs{Observations}. 
We find that analysts heavily rely on a small subset of representations for conducting their analyses (\S\ref{RESULTS-80-20}), and employ certain representations to plan and navigate subsequent steps of their analysis (\S\ref{sec:planning-aids}). We also observe a shift in analysis content, with interaction drawing analysts towards relationship observations (\S\ref{attribute-addition}). Additionally, we investigate the analysts who achieve the broadest coverage in their EDAs and describe the analysis strategies they employed to do so (\S\ref{broad-exploration}).

\subsection{Temporality, Diversity, and Velocity}
\label{representation-overview}

\begin{figure}[t!]
  \centering
  \includegraphics[width=\linewidth,alt={On the left, a faceted area chart shows the frequency of different output types over time during analysis sessions. Python code usage is prominent at the beginning and end, while visualizations dominate the middle portion. Other outputs like errors, data tables, and summary statistics occur less frequently throughout. On the right, a heatmap illustrates the relationship between visualization types and utterance categories. It reveals that scatterplots are frequently used for relationship utterances, profiles for dataset and variable utterances, and pairplots contribute across all utterance types. The heatmap also shows varying usage patterns for other chart types like barplots, dashboards, and stripplots across different utterance categories.}]{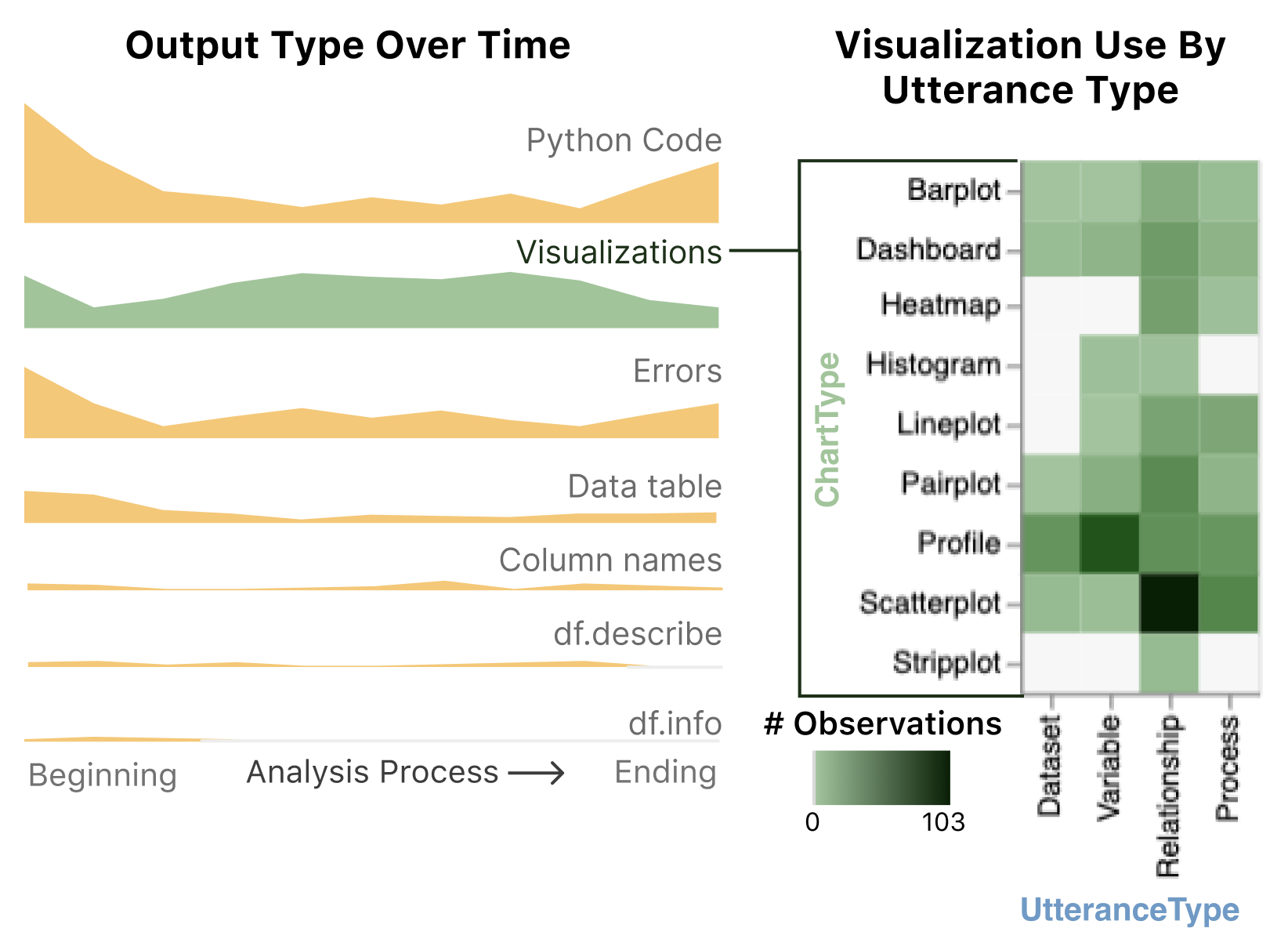}
  \caption{(left) The count of representation created over time. (right) A heatmap of the number of times different \fvis{Visualizations} were used to make an \fobs{Observation}, according to \fobs{UtteranceType}.}
  \label{fig:output_creation_and_use}
\vspace{-1em}

\end{figure}

Across all \fall{Sessions} our participants constructed a total of 1169 \foutput{Outputs}, with an individual analyst averaging 44 outputs per analysis. Python code executions were most common, especially at the beginning and end of sessions, typically for checks on central tendencies. Visualizations began to dominate about 15\% into each session, becoming the foundation for most subsequent observations (Fig.~\ref{fig:output_creation_and_use}~(left)). 
Based on this data, we introduced two metrics: \fmetric{representationDiversity}, the count of unique representations constructed during a session, and \fmetric{representationVelocity}, measuring the rate at which these representations were created. 
As Figure~\ref{fig:output_creation_and_use} shows, these metrics are moderately correlated (Pearson's $r= 0.47$); we discuss their role within analysis sessions in a subsequent section (\S~\ref{broad-exploration}). 

\begin{figure}[t!]
  \centering
  \includegraphics[width=\linewidth,alt={This image consists of two visualizations providing insights into analysis patterns during exploratory data analysis sessions. On the left, a scatterplot displays representationDiversity (number of unique types) on the x-axis and representationalVelocity (representations per minute) on the y-axis for each participant's analysis session. Notable points are labeled, such as "P3 Happiness" and "P7 Space", showing variations in analysis styles across participants and datasets. On the right, a jittered strip plot illustrates the average revisitCount for different types of representations. The plot distinguishes between all-attribute representations (darker green) and other representations (lighter green). Representations like "info", "profile", "pairplot", and "heatmap" have higher revisit counts, suggesting they are frequently used for planning and decision-making during analysis. In contrast, representations with lower revisit counts are typically used for one-off question-answering. }]{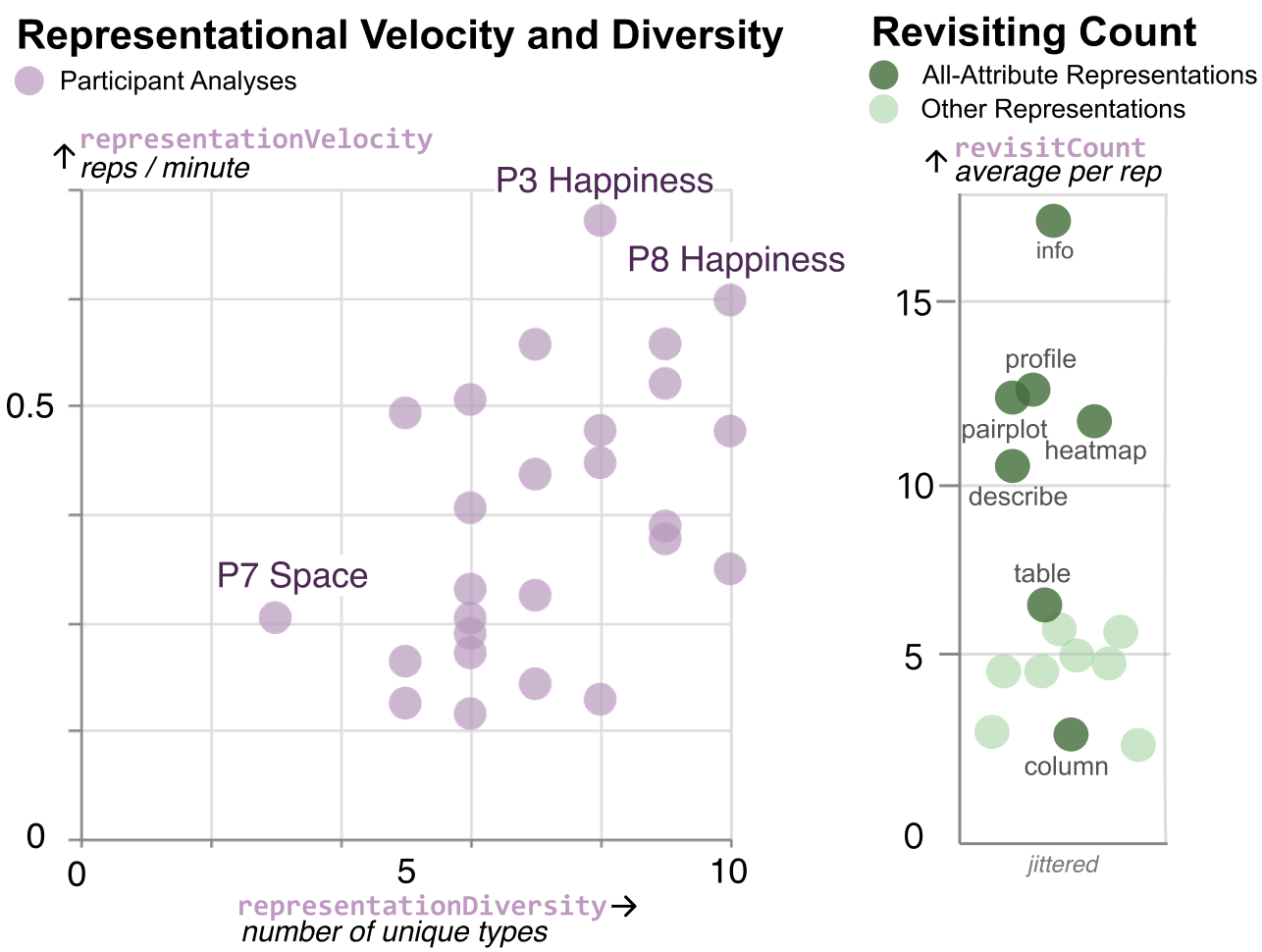}
  \caption{(left) A scatterplot of \fmetric{representationDiversity} and \fmetric{representationalVelocity} for each analysis session (\S~\ref{broad-exploration}). (right) A jittered strip plot showing average \fmetric{revisitCount} and count of \texttt{Plan of Action} utterances by Representation. Representations are colored by whether or not it is an \textit{all-attribute representation}. Representations to the bottom are typically one-off question-answering tools whereas representations to the top are frequently revisited when deciding analysis paths (\S~\ref{sec:hover-telemetry})}
  \label{fig:metrics}
  \vspace{-2em}
\end{figure}

Our analysis of the intersection of \fvis{ChartTypes} and \fobs{Observations} (Fig.~\ref{fig:output_creation_and_use}~(right)) reveals both expected and surprising usage patterns. For example, unsurprisingly, scatterplots frequently facilitated \fobs{Relationships} utterances, while profile views were used in making \fobs{Variable} utterances.
However, as Figure \ref{fig:output_creation_and_use} shows, participants would frequently use charts beyond their intended purposes or in ways that break with best practice.
For instance, \fobs{Variable} utterances constituted only 42\% of observations made with profile views\,---\,even though, ostensibly, this is the core purpose of a columnar distribution of data values.
Similarly, in contrast to visualization theory and recommender systems, which emphasize perceptual effectiveness, participant P9, a data science instructor, specifically created a representation she called a \textit{``spaghetti plot''}\,---\,a line chart with 180 different series overplotted.
Ahead of creating the chart she commented \textit{``It's going to be a bad idea''}, but persisted precisely because she wanted to ensure that the plot itself \textit{was ineffective}, as a gut check. 

\vspace{-0.25em}

\subsection{Hover Patterns and Observations}
\label{sec:hover-telemetry}

Hover patterns, captured through per-representation metrics such as \fmetric{revisitCount} and \fmetric{hoverTime}, indicate the frequency and duration of analysts' engagement with different representations. These metrics help uncover aspects of visualization usage and attention distribution that are not apparent from code execution histories alone. We combine these metrics with the \fobs{Observations} analysts made to reveal how telemetry correlates with analysis behavior. 

\vspace{-0.5em}


\subsubsection{The '80-20 Rule': Why Some Visualizations Matter More}
\label{RESULTS-80-20}
Our analysis reveals a 80-20 pattern in how participants use representations during EDA. The top 20\% of most frequently hovered representations (\textit{top-20}) accounted for 79\% of total \fmetric{hoverTime} and 75\% of observations. Representations in the \textit{top-20} had hover durations of at least 30 seconds and  an average of 2.8 Observations each, indicating deep engagement. In contrast, the bottom 80\% of representations (\textit{bottom-80}) saw significantly less use, with an average of just 0.2 observations per representation. We identify two key differences between these two sets that sheds light on analyst preferences: the ability to encode multiple attributes simultaneously, and the role of interactivity.

Representations displaying information about multiple variables simultaneously (e.g., profiles, correlation heatmaps, pairplots) were more common within the \textit{top-20}. These \textit{all-attribute representations} made up only 2\% of the \textit{bottom-80} but constituted 22\% of the \textit{top-20}, an 11-fold increase. Analysts frequently engaged with these visualizations a ``touring'' process,  previously described in \S~\ref{RESULTS-Sequential-transitions}. This involved systematically exploring the visualizations and commenting on different variable combinations approximately every 5-15 seconds. The prominence of this behavior is reflected in the extended average hover times for all-attribute visualizations, with profiles at 67 seconds, heatmaps at 75 seconds, and pairplots at 169 seconds. In contrast, we see a marked decrease in \fmetric{hoverTime} with \texttt{Code Cells} used for quick statistical checks (from 48\% of the \textit{bottom-80} to 9\% of \textit{top-20}, averaging 4.9 seconds of hovering per representation). 

Interactive visualizations were more prevalent within the \textit{top-20} (24\% of the \textit{top-20} vs. 16\% of the \textit{bottom-80}). Analysts particularly favored the \texttt{highlight\_brush} as it enabled cross-linking data subsets across multiple charts. This technique was used in over 56\% of interactive representations in the \textit{top-20}, compared to 37\% in the \textit{bottom-80}.
Similarly, the \texttt{filter\_brush} technique, which filters out all non-selected data marks from view, was used in 30\% of the interactive scatterplots found within the \textit{bottom-80}. However, \texttt{filter\_brush} went to 2\% in the \textit{top-20}, a likely side effect of filtering obscuring important context in standalone charts.

Finally, \texttt{pan\_zoom} interactions were prevalent in the \textit{bottom-80} (31\% of interactive representations) but declined to 18\% in the \textit{top-20}. Analysts consistently struggled to find effective use for pan-zoom interactions, suggesting a lack of intuition for its analytical value. Out of the 16 instances in which pan-zoom was used, we observed only one instance where it successfully uncovered an insight that would have been difficult to obtain otherwise. In this case, participant P10 zoomed into a dense, overplotted region of a scatterplot to gain more resolution, and was able to reveal a pattern in the depicted data. However, even this success story was marred by discomfort\,---\,P10 added pan-zoom to a set of horizontally arranged scatterplots that shared a common y-axis; thus, the coordinated scrolling of all scatterplots made him feel disoriented, prompting him to request \textit{``can we turn that off?''}


\subsubsection{All-Attribute Visualizations Aid Planning}
\label{sec:planning-aids}

Representations with high \fmetric{revisitCounts} (over 10 times) often serve as process planning tools, helping analysts orient themselves and prepare their next actions (Fig.~\ref{fig:metrics} (right)). A prime example of this is participant P6's use of a correlation heatmap. She created this visualization to identify the most strongly correlated attributes within her dataset and frequently returned to it as a guide for selecting specific attributes for further investigation. As she noted, \textit{``let's look at the one that is most positively correlated, which seems to be log GDP per capita. So I'll start with that variable''}. This led her to further investigate highly correlated variable sets through custom dashboards for deeper exploration, ultimately leading to an exceptional 23 observations (\S~\ref{broad-exploration}). Notably, heatmaps appeared to be particularly effective in this role, averaging 3 times as many \texttt{Process} utterances as other representations.

Such action-planning is not restricted to only visual all-attribute representations\,---\,participants frequently revisited data frame outputs (including \texttt{df.describe}, \texttt{df.info}, and the tabular output) to formulate their plans.
For instance, P11 read through the individual values of a dataframe printout, commenting: \textit{``Of course, we cannot say for the whole thing [based on just the shown rows]. So my strategy will be like going through each of the variables here, and do the summary statistic.''} 
Looking across all \fobs{Observations} tuples in our dataset, all-attribute representations are associated with \texttt{Plan of Action} utterances at a rate of 5 times higher than other representations.


\subsection{An Interactive Draw Towards Complexity}
\label{attribute-addition}

\begin{figure}[t!]
  \centering
  \includegraphics[width=\linewidth,alt={This image compares how the number of variables and data types per utterance shift between interactive and static representations in exploratory data analysis. The left side shows four bar charts: two for static and interactive profilers, and two for static and interactive scatterplots. These charts display the number of utterances for different counts of variables per utterance. The static profiler shows a high count for single-variable utterances, while the interactive profiler shows more multi-variable utterances. Similarly, the interactive scatterplot shows a shift towards more variables per utterance compared to the static scatterplot. On the right, a slope graph illustrates the change in percentage of data observations between static and interactive visualizations for different types of relationships and distributions. It shows increases in continuous-continuous relationships, multivariate relationships, and continuous distributions when using interactive visualizations, while categorical-categorical relationships decrease significantly. This visualization highlights how interactivity in data representations influences the complexity and types of observations made during analysis.}]{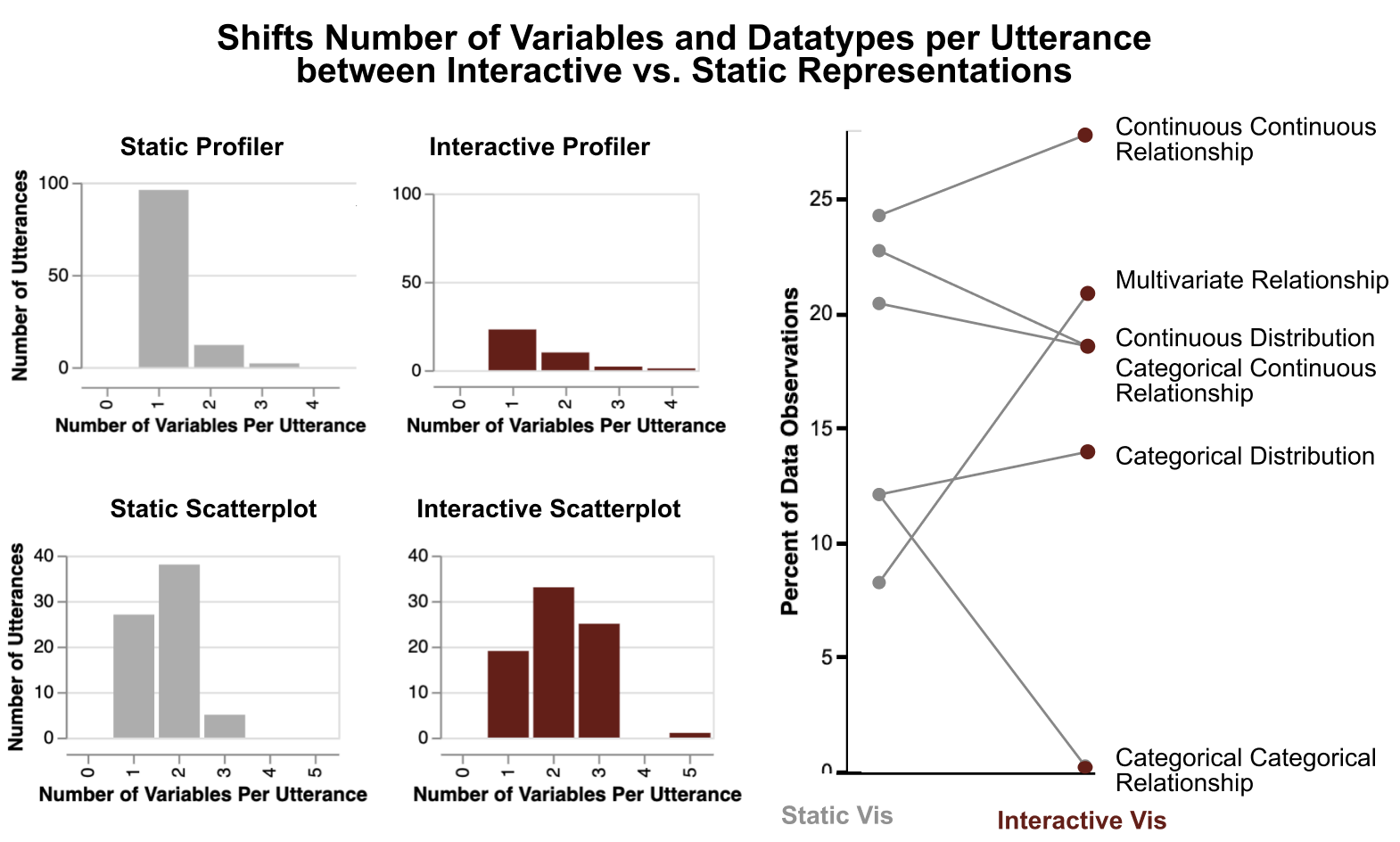}
  \caption{(left) A barchart showing the number of utterances per attribute count, faceted by whether the utterance was made using static or interactive profiler and scatterplot visualizations. 
  (right) A slope chart comparing utterance type counts between static and interactive visualizations. 
  }
  \label{fig:attribute-addition}
\vspace{-4ex}

\end{figure}

We observed correlations between the use of interactive visualizations and changes in the types and number of attributes analysts considered. When using interactive visualizations, an \textit{attribute addition} pattern emerged, where analysts' explorations moved from univariate distributions to bivariate relationships or multivariate analyses.
For example, participant P6 used a static profile visualization to analyze the univariate distributions of her columns, making 6 utterances about their distributions. At the beginning of the interactive session, she created an interactive version of the profile, and immediately began using it to analyze relationships\,---\,brushing on the chart to examine a target population and generating 6 new utterances about that population's relationship to other variables. This pattern of behavior persisted across datasets for other participants (Fig.~\ref{fig:attribute-addition}~(left)). Analysts consistently leveraged interactivity to deepen their exploration, sometimes even skipping over distributional analyses to instead analyze more complex data relationships (\S~\ref{sec:variable-gap}).

We also observed shifts in behavior prompted by filtering interactions in scatterplots (Fig.~\ref{fig:attribute-addition}~(right)). Prior to the interactive session, we observed participants discussing bivariate relationships using scatterplots; however, when interaction was added, their utterances tended to focus on the multivariate relationships. Multiple participants used brushes to extract subsets from data clusters and pursued analysis paths to differentiate that cluster from the rest of the data. Another case of this was the use of the \texttt{filter\_slider}, an interaction technique which filters the chart to only the data value present in a particular value on a slider query widget. The shift we observe between these interactive and static charts presents the allure of interactive representations, seemingly pulling analysts towards investigating more complicated relationships even when those interactions are not actively being used. 

However, attribute addition behavior was not observed equally across data types. Our participants often used interactive visualizations for multivariate (frequently all continuous variables) and continuous x continuous bivariate relationships (Fig.~\ref{fig:attribute-addition}~(right)). However we note the overall patterns are most salient at the aggregate level and the participant level contains sparsity in the utterances made for a given data type. Thus while we chose to report the results to fully describe the behavior that we saw, such descriptions warrant additional investigations to understand the role that interaction may play in drawing analyst hypotheses towards more multivariate and complex relationships and if such patterns exist during longer EDA sessions.

\subsection{Patterns of Broad Observation Space Exploration}
\label{broad-exploration}

\begin{figure}[t!]
  \centering
  \includegraphics[width=\linewidth,alt={The top part shows a stripplot displaying the number of unique observations per analysis session for three categories: Dataset, Variable, and Relationship. Each point represents a session, color-coded for interactive (maroon) or static (gray) conditions. The plot reveals variations in observation counts across different analysis types and conditions, with some sessions marked for their distinctive patterns (P8, P3, P6). The bottom part features three heatmaps representing attribute co-occurrences in relationship observations for participants P8, P3, and P6. Each heatmap shows how frequently different variables were explored together, with darker colors indicating higher co-occurrence. The heatmaps are labeled with different search strategies: Parameterized Search (P8), Iterative Deepening (P3), and Best First Search (P6), illustrating distinct approaches to data exploration. This visualization allows for comparison of analysis depth and patterns between interactive and static conditions, as well as individual differences in exploration strategies among participants.}]{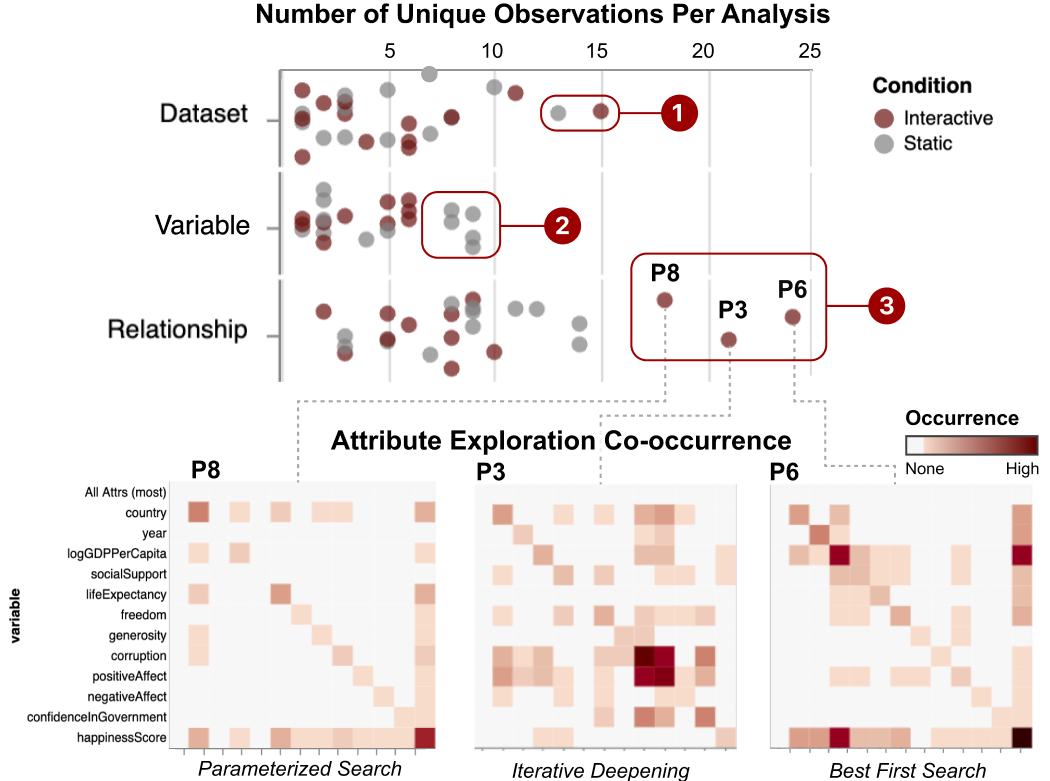}
  \caption{(top) A stripplot of percent of total unique \fobs{Observations} visited per analysis session, broken down by high level type and colored by Analysis Condition (Interactive or Static). (bottom) Heatmaps representing attribute co-occurrences when participants made observations about relationships between variables.
  }
  \vspace{-2em}
  \label{fig:broad-states}
\end{figure}

Previous studies have characterized EDAs based on the number of attributes analysts considered~\cite{battle_characterizing_2019, wongsuphasawat_voyager_2017}. We build on this approach, applying it to our more comprehensive definition of \fobs{Observations}, which encompasses both what was learned (\fobs{UtteranceType}) and which data \fobs{Attributes} were considered. 
Adapting Battle et al.'s method~\cite{battle_characterizing_2019}, we created binary histograms representing whether participants made a specific utterance type on an attribute set (e.g., observed the \textit{relationship} between \textit{happiness} and \textit{GDP}). By calculating the percentage of total possible states each participant explored, we can rank participants by their breadth of exploration and investigate the ways in which \foutput{Representations} changed the analysis \fall{Session}. For example, participant P9, a data science instructor, made the most extensive \fobs{Dataset} observations across both static and interactive conditions (Fig.~\ref{fig:broad-states}~\circledannotation{1}). These observations occurred as P9 began each of her analysis sessions with a variable metadata tour: systematically going through each attribute in the data dictionary, spending time discussing what the variable meant and her opinions on its usefulness. 
Similarly, we observe the 5 participants who made the most \fobs{Variable} utterances (Fig.~\ref{fig:broad-states}~~\circledannotation{2}) did so in the static condition using profile visualizations.


In contrast, approaches for exploring a broad set of \fobs{Relationship} observations (Fig.~\ref{fig:broad-states}~\circledannotation{3}) reveals a diverse set of strategies. To investigate these patterns of exploration, we created attribute co-occurance heatmaps (Fig.~\ref{fig:broad-states}) to ``fingerprint''  and explain these strategies: 

\textbf{P8: Parameterized Search.} Driven by a clear goal and an aversion to \textit{``mindless''} exploration, P8 adopted a systematic, iterative approach reminiscent of a parameterized search through \foutput{Representations} and \fvis{Encodings}. She cycled through which attributes were mapped to encodings (e.g., \texttt{scatterplot(y=happiness, x=column[index])}), methodically investigating potential relationships between each attribute and the outcome variable. When she encountered specific patterns of interest, she then modified her scatterplot, adding interactions such as brushes and tooltips to investigate outliers and subsets. The resultant fingerprint visualization depicts a focused analysis centered on the outcome variable, with some targeted off-diagonal probes into the country, investigated using tooltips and brushes.  

\textbf{P3: Iterative Deepening.} P3's approach was guided by emergent patterns in the data, resembling an iterative deepening search.
He generated scatterplots based on his intuition for interesting relationships, largely ignoring the outcome variable. This is reflected in his focus on variables other than \textit{happinessScore} (bottom row and right column). Upon noticing clusters, he investigated their characteristics, iterating through interactions and encodings (adding tooltips, brushes and \fvis{color} encodings) to identify potential explanatory variables. This behavior is captured in his high \fmetric{representationalVelocity} and \fmetric{representationDiversity} as shown in Figure ~\ref{fig:metrics}~(left), suggesting he wasn't wedded to a single visualization type but explored various options to find insights. This iterative deepening process ultimately led to a scattered thumbprint reflecting his serendipitous journey through attribute space, driven by unexpected findings. 

\textbf{P6: Heuristic-Guided Best First Search.} P6's approach combined a methodical foundation with responsive, opportunistic elements characteristic of best-first search \cite{pearl1984heuristics}. This strategy prioritizes exploring the most promising nodes within a search space based on a pre-defined heuristic. P6's analysis mirrored this approach by selecting attributes to plot based on their correlation with her outcome variable. After analyzing these attribute sets in a custom dashboard, she would return to her correlation matrix to choose her next attribute set, effectively "touring" through her correlation matrix. She revisited this matrix 35 times during her analysis, demonstrating a high \fmetric{revisitCount} for this visualization. This strategy produced a cohesive analysis that investigated both direct predictors and potential confounds of the outcome variable, evident in her targeted analysis along the bottom row and off-diagonal of her thumbprint visualization. 


\vspace{-0.5em}
\subsection{Thinking in the Language of Interaction}
\label{RESULT:language-of-interaction}
In interaction design, perceived affordances \cite{norman_design_2013} signal the operations a user believes are possible within an interface. Well-designed affordances establish interaction dynamics\,---\,the rules governing how users interact with the interface. Our study revealed that data scientists reasoned about these dynamics to generate new analytical hypotheses. In other words, they translated ``the language of interaction'' into novel analytical questions.
As participant P8 described: \textit{``My thought of intersecting High GDP and High Life-Expectancy [countries] happened precisely because there was interaction... I was thinking, 'Oh I wonder if multi-select works'... That is actually what led me to think, 'Oh this would also be interesting on an analytical level.'''}. Later she commented that such an insight \textit{``would not have occurred to me if not for the fact I was working with an interactive visualization.''} 

Participant P6's insights emerged from a similar process of experimentation.  
Having successfully used ALX's copy-and-paste technique to paste filters between charts, he began to consider the broader possibilities this interaction technique offered.
While browsing other charts, he stumbled upon a bar plot showing the count of records over time. Intrigued, he initially tested if the copy-and-paste would function in this context. However, a spark ignited: rather than a simple test of function, he realized it would be more insightful to filter on the most recent years of data. This act of guided experimentation, prompted by the affordances of an interaction design (rather than performing the interaction itself and observing any updates), led him to discover an unexpected trend in life expectancy over time.

These examples suggest that interactive features play a more generative role in analysis than typically acknowledged. While existing literature often focuses on interactions as tools for completing specific tasks, our observations reveal that the rules of the interaction design can inform emerging hypotheses and shape analytical reasoning. This insight has two key implications. First, there's an opportunity to critically examine how we articulate and implement the constraints and rules of interaction dynamics. Different designs may substantially impact how analysts reason about these rules and, consequently, how they approach their analysis. Second, beyond investigating how visual cues influence interaction usage \cite{boy_suggested_2016}, future studies should explore how various cues shape analysts' conceptualization and potential application of interaction techniques. By recognizing the interplay between interaction mechanics and analytical cognition, we can pave the way for tools that more effectively partner with the analyst during the discovery process.
\vspace{-0.75em}
\vspace{-0.5em}
\section{Discussion and Future Work}

In this paper, we conducted a qualitative experiment to richly characterize the \emph{situated} nature of EDA in computational notebooks. Through mixed-methods analysis of utterances and telemetry, we developed a formal description of EDA sessions and applied it to analyze 26 sessions by 13 data science professionals.
In response to \textbf{RQ1}, we uncovered distinct temporal patterns in analysts' \fobs{Observations}, revealing how different types of insights evolve throughout an EDA session. We identified phenomena such as \emph{attribute-addition} and \emph{reasoning in the language of interaction}, which shed light on the cognitive processes underlying EDA in computational notebooks. 
Addressing \textbf{RQ2}, our analysis uncovered substantial differences in how analysts use interactive versus static visualizations. Interactive visualizations often led to earlier discoveries of relationships between dataset attributes. Analysts also tended to rely heavily on a small subset of representations, with interactive visualizations comprising a sizeable portion of this subset. Finally, we introduce metrics such as \fmetric{revisitCount}, \fmetric{representationalDiversity}, and \fmetric{representationalVelocity} to quantify broad coverage in EDA. Our work contributes to calls for investigating the theoretical foundation of EDA \cite{hullman_designing_2021} and offers principles for designing more analyst-aligned EDA tools.

\vspace{-0.1em}
\subsection{Limitations}

Although our approach yielded useful insights about how data science professionals analyze data, we note that studying EDA in a laboratory context poses some inherent limitations.
For example, think-aloud protocols may artificially structure thought processes that are more fluid in unobserved settings (e.g., participants may prioritize tasks that are easier to articulate)~\cite{davies_cognitive_2005}. However, in comparison to post hoc reflections, thinking aloud provided \textit{in situ} insights that captured important nuance, and aligns with approaches used in other studies \cite{arias-hernandez_joint_2011}. 

Our study's sample (N=13) may not fully represent the diversity of approaches to EDA. However, this size aligns with qualitative research practices that prioritize depth over breadth \cite{daniel_using_2019}. Thematic saturation observed in our data also suggests that the identified themes provide robust insights into the EDA process.


The 25-minute time limit per analysis may have also constrained the range of analyses participants engaged in.
This time limit, consistent with prior visualization studies~\cite{wu_b2_2020, wongsuphasawat_voyager_2017, battle_characterizing_2019}, balances the need to maintain participant engagement without requiring extended time commitments.
Research shows that analysts often encounter time-sensitive tasks in their work\cite{wongsuphasawat_goals_2019}, and in practice, we did not abruptly cut participants off.
Thus, on average, participants took 29 minutes to complete an analysis. 

Finally, using a new visualization library inevitably presents challenges to analysts and may introduce novelty effects, especially for those accustomed to static visualizations. 
We sought to mitigate these effects in two ways. 
First, we allocated 20-minutes to demonstrations and tutorials of the library. 
Second, ALX was intentionally designed as a visualization and interaction \textit{typology} (as opposed to a more composable grammar) to minimize specification difficulty\,---\,with the terms of the two typologies designed to mirror common visualization and interaction design patterns.
More importantly, introducing a new library allowed us to control for participant expertise, as analysts did not have prior tool-specific habits that could have confounded our comparison of analysis sessions. These sessions, therefore, reflect a "first-use study," which is common in studies of EDA activity~\cite{wu_b2_2020, wongsuphasawat_voyager_2017, zgraggen_investigating_2018, kale_evm_2023}. 


\subsection{Implications for EDA Tool Design}

Our results suggest several opportunities for interactive visualization tooling to better support EDA. For instance, several of our participants engaged \textit{touring} to systematically explore the data (\S~\ref{RESULTS-Sequential-transitions}). Yet, existing tools provide poor support for such activity, largely leaving analysts to drive interactions based on their priors and hypotheses they may wish to answer. Akin to visualization recommender systems~\cite{lab_draco_2018}, novel EDA tooling might instead leverage nascent grammars~\cite{suh_grammar_2022} to systematically enumerate the space of hypotheses that can be interactively reached with a given visualization, and proactively suggest particular analysis paths. By leveraging information scent~\cite{willett_scented_2007}, such tools could help analysts think more deeply in the \textit{language of interaction} (\S~\ref{RESULT:language-of-interaction})\,---\,that is, even if an analyst did not adopt a suggestion for an interactive path, the suggestion itself may prompt them to think in different ways. 

Relatedly, we found our participants' use of visualizations as \textit{action planning aids (\S~\ref{sec:planning-aids})} striking. In computational notebooks, where visualizations are linearly presented, several participants were willing to pay a ``scrolling tax'' to reach these representations. While some research systems have explored mechanisms for making such representations more readily available (e.g., B2 stitches a visual analytics dashboard alongside a linear notebook~\cite{wu_b2_2020}), our results suggest a wider opportunity. For instance, although research has identified the merit of overview+detail or focus+context techniques, few visualization libraries support them out-of-the-box. When they do, these techniques are supported in relatively limited ways (e.g., when panning/zooming a scatterplot or map). Our results suggest the need for more generalized support for wayfinding\,---\,especially to coordinate multiple separate visualizations. Here, we find the \textit{interaction snapshots} ~\cite{wu_facilitating_2020} and \textit{EDA assistant} ~\cite{li_edassistant_2023} particularly promising for displaying the range of plausible next actions, and enabling quick probing of the analysis space.

The prevalance of \fobs{Process} utterances during analysis sessions illustrates that participants engage in a level of metacognition\,---\, thinking about their own thinking. How might visual analysis tools better support process reflections across visualization creation, interaction design, code, and statistical output? Drawing on research in distributed cognition~\cite{hill_edit_1992}, we envision that displays of analysis histories could foster valuable self-reflection. Systems like Lumos~\cite{narechania_lumos_2022, li_edassistant_2023} are already exploring this, highlighting a rich research space. For example, what marks a significant point in the analytical journey? While our formalism points to \fobs{Observations} and \foutput{Representation} creation as key moments, analysts may have different views when reflecting on their own activity.

\vspace{-0.35em}
\subsection{Studying Interactive Analysis as Situated Activity}

Our work was motivated by a desire to study interaction as \textit{situated activity}\,---\,that is, involving human analysts working in a particular context, externalizing their cognition through visual representations, and interactively making observations with them.
While valuable, we believe this paper takes only an initial step towards this approach.
To complement recent  work that looks to scale-up our ability to study interaction (e.g., through benchmarks~\cite{gathani_grammar-based_2022} and novel systems~\cite{dotan_track_2019, nobre_revisit_2021}), we advocate for methods that allow us to study it \textit{more closely}.

We find methods from sociolinguistics and linguistic anthropology used to analyze interpersonal interaction particularly compelling. 
For instance, discourse and conversational analysis~\cite{sacks_simplest_1974} involves a meticulous examination of conversation transcripts, and has been used by researchers to make fundamental linguistic discoveries such as turn-taking~\cite{sacks_simplest_1974}.
While visualization researchers are beginning to draw on such linguistic theories to inform interaction design guidelines~\cite{setlur2022converse, setlur_heuristics_2023}, we believe there is a ripe opportunity to adapt them for analyzing interactive behavior as well. 
For instance, the development of a specialized notation system was particularly crucial to the success of conversational analysis\,---\,allowing researchers to annotate linguistic features such as prosody, tone, pitch, pauses, and gaze. 
What would an equivalent notation for analyzing interaction look like? 
Similarly, systems for conversational analysis enable flexible definitions of analytic units and abstractions.
In contrast, existing interaction provenance systems~\cite{lex_opportunities_2021} largely follow a dichotomy of either low-level event logs (e.g., mouse movements, clicks, etc.) or high-level semantically meaningful events (e.g., filter, explore, etc.)\,---\,future systems must grapple with how to support more fluid analysis between these levels.
Finally, as our study demonstrates, to ``closely read'' interactive behavior requires capturing a rich multimodal data streams.
Simply concatenating and visually linking these streams together risks introducing ambiguities in understand the precise sequences and potential causal relationships between measures. 
Rather, akin to systems like ChronoViz~\cite{fouse_chronoviz_2011}, we envision future systems offering richer juxtapositions of this multimodal data.

\section{Acknowledgements}
\label{Acknowledgements}
We thank our study participants and anonymous reviewers for their thoughtful comments. This work was supported by NSF grants \#1942659 and \#1900991. This material is based upon work supported by the National Science Foundation under Grant No. 2141064.



\bibliographystyle{abbrv-doi}

\bibliography{template}

\end{document}